\documentclass{JHEP3}

\title{Rarita-Schwinger  equation  from  principle equation for all spins }
\author{B. Sazdovi\'c \\
        Institute of Physics, P.O .Box 57, 11001 Belgrade, Serbia \\
        E-mail: \email{sazdovic@phy.bg.ac.rs}}

\abstract{

In previous two articles we postulated that field equations for arbitrary spin and helicity are Casimir eigenvalue equations.  In massive case, from such principle equation,
we derived  spin-$0$ Klein-Gordon,  spin-$\frac{1}{2}$  Dirac  and spin-$1$  vector  equations. In the present article we will derived spin-$\frac{3}{2}$  Rarita-Schwinger equation, which is nontrivial combination of vector and spinor case.  We will  also show that vector-spinor field  contains two spin-$\frac{1}{2}$ Dirac fields.

  }

\preprint{}

\usepackage{amsmath}

\begin{document}

\section{Introduction}
\setcounter{equation}{0}

From a physical point of view equations of motion for free fields are  differential equations which separate irreducible representation of Poincare group \cite{Wi, BW, W}.
From a mathematical  point of view Casimir invariants label irreducible representations of the group.
So, it is natural to expect that  Casimir invariants (in fact theirs  eigenvalues)   define the characteristics of the particles. The  number of characteristics is equal to number of Casimir invariants for given group.
In the case of massive Poincare group there are two Casimir operators and consequently there are two characteristics of the particles,  mass $m$  and   spin  $s$.
In the case of massless  Poincare group there is one Casimir operator and consequently there is one  characteristics of the particles, helicity.

In  previous two articles  \cite{BS1, BS2} we introduced  principle equations  which are source of  all known massive and massless field equations for arbitrary spin and helicity.
They are  eigenvalue equations  for Casimir operators of Poincare group.

For massless particles in Ref.\cite{BS2}  we have presented  examples of  representations  with helicities $\lambda = \pm 1$  (vector field, symmetric and antisymmetric   second rang  tensor)
which  describe electromagnetic interaction  and   representations  with helicities $\lambda = \pm 2$ (symmetric second rang tensor and fourth rang tensor with appropriate symmetries)
which  describe  gravitation  interaction.

For massive  particles in Ref.\cite{BS1}  we have presented  examples of scalar field, Dirac spinor and vector field  with spins $0$, $\frac{1}{2}$  and $1$.
Here we  will continue this approach  and show that spin-$\frac{3}{2}$  particles which is described with  Rarita-Schwinger equation  \cite{RS}  follows  from the same massive principle equation.

This confirms the main idea that all massive field theory equations for arbitrary spin can be describe by two principle equations  (because  in that case  there are two Casimir operators)
and  all field theory equations for arbitrary  helicity  can be describe by one  principle equation  (because  in that case there is one Casimir operator).

To simplify expressions,  in some parts of the  article,  when it does not cause confusion we will omit spinor indices  and sometimes  both vector and spinor indices.

\section{Massive principle field equations}
\setcounter{equation}{0}

In this section we will repeat briefly  the main parts  of Ref.\cite{BS1}   which we will need for further presentation.

\subsection{Representations of  the Poincare  algebra and  Casimir operators}

Let us  introduce $P_a$ as translation generators and  $ M_{a b}   =   L_{a b} +   S_{a b}$  as four dimensional  rotations generators where    $L_{a b} =    x_a P_b - x_b P_a $   is  orbital part   and  $S_{a b}$  is spin part.
Then Poincare algebra  has the  form
\begin{eqnarray}\label{eq:PG}
[P_a, P_b] = 0  \, , \qquad  [P_c, M_{a b}]  =   - i \Big( \eta_{b c} P_a  - \eta_{a c } P_b  \Big)   \, ,  \qquad
\end{eqnarray}
\begin{eqnarray}\label{eq:}
[M_{a b}, M_{c d}]  =  i   \Big( \eta_{a d} M_{b c} + \eta_{b c} M_{a d} - \eta_{a c} M_{b d} - \eta_{b d} M_{a c} \Big)   \,  .
\end{eqnarray}

Casimir operators  commute  with all  group  generators  and  allow us to label the irreducible representations of the group.
For   $P^2 > 0$  there are   two Casimir operators
\begin{eqnarray}\label{eq:COP}
P^2 = P^a P_a \, ,  \qquad     W^2  =  - \frac{P^2}{2}  M_{a b}  M^{a b}  \, ,
\end{eqnarray}
where
\begin{eqnarray}\label{eq:PLv}
W_a = \frac{1}{2} \varepsilon_{a b c d} M^{b c} P^d     \, ,
\end{eqnarray}
is  Pauli-Lubanski vector.  It does not depend  on orbital part  $L_{a b} $,   because  $ \varepsilon_{a b c d} L^{b c} P^d =    0 $.
Therefore,  from Casimir  operator's point of view  only   generators $P_a$  and $S_{a b}$   are relevant  and we can rewrite above equations in the form
\begin{eqnarray}\label{eq:COP1}
P^2 = P^a P_a \, ,  \qquad     W^2  =  - \frac{P^2}{2}  S_{a b}  S^{a b}  \, ,   \qquad  W_a = \frac{1}{2} \varepsilon_{a b c d} S^{b c} P^d     \, .
\end{eqnarray}
They satisfy  commutation  relations
\begin{eqnarray}\label{eq:PGS}
[P_a, P_b] = 0  \, , \qquad  [P_c, S_{a b}]  =  0  \, ,
\end{eqnarray}
\begin{eqnarray}\label{eq:PGSS}
[S_{a b} , S_{c d}]  =  i   \Big( \eta_{a d} S_{b c} + \eta_{b c} S_{a d} - \eta_{a c} S_{b d} - \eta_{b d} S_{a c} \Big)   \,  .
\end{eqnarray}
Note that  now  generators  $P_a$ and  $S_{a b}$ commute  so that  Casimir  operators   are defined  without problem of order  unambiguity.

Representations of Poincare group  are labeled by the  eigenvalues of Casimir invariants,   that we can assign to a physical state.
For $P^2 > 0$ eigenvalues of Casimir operators are    mass $m$  and   spin  $s$
\begin{eqnarray}\label{eq:EvCo}
P^2 = m^2 \, ,  \qquad  W^2 =   - m^2   s(s+1) \, .
\end{eqnarray}

We will work  with arbitrary field   $\Psi^A (x)$ which transforms with corresponding  representation of Poincare group,   where $A$ contains  set of vector and spinor indices.

To find  representation of Casimir operators we need representation of Poincare algebra  generators,  momentum  $P_a$  and spin   $ S_{a b}$.
Representation of momentum  $P_a$ is well known from quantum mechanics $(P_a)^A{}_B \to i   \delta^A_B  \partial_a$ and it is  spins independent.

We can  obtain   representation of spin generators  $ (S_{a b})^A{}_B$  for  arbitrary field   from corresponding expressions with smaller spins and initial expression for fermions.
As shown in Ref.\cite{BS1}   spin generators   $(S_{a b})^A{}_B$ act as derivatives
\begin{eqnarray}\label{eq:BEder}
 (S_{a b})^{A B}{}_{C D}    =  (S_{a b})^{A}{}_{C} \delta_D^B  +  \delta^A_C (S_{a b})^{B}{}_{D}  \, .
\end{eqnarray}
It is easy to check that   $(S_{a b})^{A B}{}_{C D} $  is solution of  (\ref{eq:PGSS}).

In the present article we will need the  initial expressions for representation of spin generators  for   Dirac spinor   and vector fields
\begin{eqnarray}\label{eq:gLr21}
(S_{a b})^\alpha{}_\beta   \to  (f_{a b})^\alpha{}_\beta = \frac{i}{4} [\gamma_a, \gamma_b ]^\alpha{}_\beta  \, ,  \qquad
 (S_{a b})^c{}_d  \to   (s_{a b})^c{}_d  = i \Big(    \delta^c_a   \eta_{b d} - \delta^c_b   \eta_{a d}  \Big)   \,  .
\end{eqnarray}
For  later convenience,  we will keep notation  $(S_{a b})^{a \alpha}{}_{b \beta}$ for spin operator of  Rarite-Schwinger field  and  introduced  notation
$(f_{a b})^\alpha{}_\beta $ and   $(s_{a b})^c{}_d $  for    Dirac   and vector field.

Note that with the help of  these expressions  we  can  find  representations  for  fields  with arbitrary spin   using recurrence relation (\ref{eq:BEder}).
In the present article we will   use  expressions  (\ref{eq:BEder})  and   (\ref{eq:gLr21})  to find  representations  for spin-$\frac{3}{2}$ fields.

\subsection{Principle field equations}

In Ref.\cite{BS1}   we  postulated  principle field equations for  arbitrary  spin as   representation   of  relations  (\ref{eq:EvCo})
\begin{eqnarray}\label{eq:BS}
(P^2)^A{}_B  \Psi^B (x)  =  m   \Psi^A (x)  \, ,  \qquad   {\cal S}^A{}_B    \Psi^B (x) =  s (s + 1)  \Psi^A (x)   \, ,
\end{eqnarray}
where   $(P^2)^A{}_B$ and ${\cal S}^A{}_B $ are representation of Casimir operators.  Theirs  eigenvalues   $m$ and   $s (s + 1)$   define the mass $m$  and  the  spins $s$.

These equations  are  Poincare covariant because Casimir operators commute with all Poincare generators.
In particular,  Casimir operators  commute mutually  and they   have  a complete set of common eigenfunctions.  So,  we are able to impose both   Casimir operators to the same field $\Psi^A (x)$.

Using representations for  momentum and spin operators  we obtain the  set of differential  equations
\begin{eqnarray}\label{eq:BS1}
\Big( \partial^2 + m^2 \Big)  \Psi^A (x)  = 0  \, ,  \qquad {\cal S}^A{}_B    \Psi^B (x) =  s (s + 1)  \Psi^A (x)   \, ,
\end{eqnarray}
where
\begin{eqnarray}\label{eq:Oper}
{\cal S}^A{}_B  \equiv - \frac{(W^2)^A{}_B}{m^2} =  - \frac{1}{m^2}     (S^a{}_c)^A{}_C  (S^{c b})^C{}_B \partial_a \partial_b   +  \frac{1}{2}   (S_{a b})^A{}_C  (S^{a b})^C{}_B     \, .
\end{eqnarray}
In momentum space we have
\begin{eqnarray}\label{eq:BSm}
\Big( p^2 - m^2 \Big)  \Psi^A (p)  = 0  \, ,  \qquad ({\cal S}_p)^A{}_B    \Psi^B (p) =  s (s + 1)  \Psi^A (p)   \, ,
\end{eqnarray}
where
\begin{eqnarray}\label{eq:Operm}
({\cal S}_p)^A{}_B  \equiv - \frac{(W^2)^A{}_B}{m} =   \frac{1}{m^2}     (S^a{}_c)^A{}_C  (S^{c b})^C{}_B p_a p_b   +  \frac{1}{2}   (S_{a b})^A{}_C  (S^{a b})^C{}_B     \, .
\end{eqnarray}

\subsection{Principle  field equations  for standard momentum}

Starting from  principle field   equations  we  are able  to obtain particular equations for any given spin.
In the present  article from vector-spinor field  $\psi^{a \alpha}$  we will separate  two  spin-$\frac{1}{2}$ fields  and one   spin-$\frac{3}{2}$ field.
We will show that    spin-$\frac{3}{2}$ field   satisfy   Rarita-Schwinger  equation  and  two spin-$\frac{1}{2}$ fields   satisfy  Dirac   equations.
In order to  to achieve that  it is useful to go to standard momentum.

For massive case where  $p^2 = m^2$  we can chose   rest frame  momentum   $k^a = (m, 0, 0, 0  )$  as standard momentum.
With this choice the first equation  (\ref{eq:BSm})  is solved.  We can  express arbitrary  momentum  $p^a$  as Lorentz transformation of $k^a$
\begin{eqnarray}\label{eq:paLka}
p^a =  L^a{}_b  (p)  k^b \, .
\end{eqnarray}

For standard momentum second  differential equation (\ref{eq:BSm})   becomes  algebraic one.
After solving  algebraic equation we can go back to  $p^a$ dependent solutions using  (\ref{eq:paLka}) and then to solution in coordinate representation.

The spin  equation  in the frame of standard momentum   takes the  form
\begin{eqnarray}\label{eq:BSms}
{\cal S}^A{}_B    \Psi^B (k) =  s (s + 1)  \Psi^A (k)   \, ,
\end{eqnarray}
where
\begin{eqnarray}\label{eq:Operr1}
{\cal S}^A{}_B  =     (S^2_i)^A{}_B    \, ,  \qquad  (S_i)^A{}_B    = \frac{1}{2} \, \varepsilon_{i j k} (S_{j k} )^A{}_B  \, .
\end{eqnarray}

Note that instead of  $6$ components of spin operator $S_{a b}$ in the frame of standard  momentum   we have left with $3$  components  $S_i$,  which are generators of space rotations.
They form a subgroup known as {\it little group} for massive Poinacare case.

\subsection{Massive field equation for selected spin }

The next step is to  solve the eigenvalue problem of operator  ${\cal S}^A{}_B$  (\ref{eq:BSms}).
To have nontrivial solution for function $\Psi^A $ the    characteristic   polynomial   must vanish
\begin{eqnarray}\label{eq:DetSs}
  \det \Big(  {\cal S}^A{}_B  -   \lambda \delta^A_B   \Big)  = 0      \, ,         \qquad   \lambda   \equiv   s ( s + 1)  \, .
\end{eqnarray}
It  may have multiple solutions, $\lambda_i$  where  $i = 1, 2, \cdots , n$  counts different   representations.
The values $s_i$,  corresponding  to the  eigenvalues $\lambda_i$,   are spins of   different  representations.

The   representations  of eigenfunctions  $\Psi_i^A (k)$  with selected  spin  $s_i$  have a form
 \begin{eqnarray}\label{eq:IrrPa}
\Psi_i^A (k)  = ( P_i )^A{}_B \Psi^B  (k) \, ,     \qquad
( P_i )^A{}_B =   \frac{  \Big[  \prod_{j \neq i}^n    \Big(  {\cal S}   - \lambda_j    \Big) \Big]^A{}_B }{  \prod_{j \neq i}^n    \Big(  \lambda_i   - \lambda_j  \Big) }        \, ,  \qquad
 i =  \{ 1, 2, \cdots , n  \} \, ,    \qquad
\end{eqnarray}
where $(P_i)^A{}_B$ are corresponding  projection operators. To  obtain   irreducible representations we should  separate fields $\Psi_i^A  (k)$ in several  sets  with well-defined symmetry properties.

We can rewrite equation   (\ref{eq:DetSs})   in the form
\begin{eqnarray}\label{eq:Smlade}
\det \Big(  {\cal S}^A{}_B  -   \lambda \delta^A_B   \Big)
=     \prod_{i =0}^n   ( \lambda_i  - \lambda )^{d_i}  P_i = 0         \,   ,    \qquad
\end{eqnarray}
where  $d_i$ are  dimensions of eigenpaces which correspond   to projector  $P_i$.
As shown in Ref.\cite{BS1, BS2}     $d_i$  is  number of degrees of freedom of component   $\Psi_i^A  (k)$.

In general case it is not easy to find straightforward solutions of equation  (\ref{eq:DetSs}).  For example, in Rarita-Schwinger case we need to calculate   determinant $16 \times 16$.
To avoid this complicated calculation we have developed the projector method which makes the calculation much easier.
In this  method projectors $( P_i )^{a \alpha}{}_{b \beta}$ of Rarita-Schwinger field are  combination of elementary projectors of Dirac and vector fields.

Solution $\Psi_i^A (k)$  can satisfy  some   supplementary condition.  In fact all projection operators $P_j \, , (j \ne i ) $  annihilate $\Psi_i^A$,
so  that  supplementary  conditions   take the form
\begin{eqnarray}\label{eq:SCgen}
( P_j )^A{}_B  (k) \Psi_i^B  (k)  =  0  \, ,  \qquad  (j \ne i)    \, .
\end{eqnarray}

After finding  solution for standard momentum    $\Psi_i^A (k)$,   we can  boost  it  to the arbitrary frame
$\Psi_i^A  (k) \to \Psi_i^A (p)   = {\cal B}^A{}_B (\mathbf{p})  \Psi_i^B (k)$  and  obtain solution of the  second equations  (\ref{eq:BSm}). With the help of relation $p_a  \to  i \partial_a $
we can find  coordinate representation  of function  $\Psi_i^A  (x)$.

The frame of standard momentum solves Klein-Gordon equation, but in arbitrary frame we will have to take it into account.  In coordinate representation it has a form
\begin{eqnarray}\label{eq:Dlcon}
 (\partial^2 +  m^2   )  (P_i)^A{}_B  \Psi_i^B  (x) = 0  \, ,  \qquad
\end{eqnarray}
So,   equations (\ref{eq:SCgen})  and  (\ref{eq:Dlcon})  are complete set of   conditions which  solution  $\Psi_i^A  (x)$   should  satisfy.

We are looking for an equation   whose solution is  $\Psi_i^A  (x)$.  It must satisfy the same conditions  and  consequently  consists of
Klein-Gordon equation  plus supplementary conditions
\begin{eqnarray}\label{eq:}
 (\partial^2 +  m^2   )   (P_i)^A{}_B \Psi^B  (x) = 0  \, ,  \qquad    ( P_j )^A{}_B    \Psi^B  (x)  =  0  \, .  \qquad   (j \ne i  )
\end{eqnarray}

These   equations  can be combined  into  one equation  as linear combination of  Klein-Gordon    equation and supplementary conditions
\begin{eqnarray}\label{eq:coN1IKG}
\Big[  ( \partial^2 +  m^2  ) P_i  +  \sum_{j \ne i}   C_j P_j    \Big]^A{}_B  \Psi^B  (x)  =  0    \,  .  \qquad
\end{eqnarray}

If index $A$ contain  spinor index $\alpha$, which means that $A = \{\alpha, \tilde{A} \}$  then  we can   linearize Klein-Gordon  equation and  obtain first order Dirac like  equation
\begin{eqnarray}\label{eq:DllKG}
 (i \hat{\partial}  - m  )^\alpha{}_\beta   \Psi_i^{\beta   \tilde{A} }  (x) = 0   \,  ,    \qquad   (  \hat{\partial}  =  \gamma^a \partial_a )    \, ,
\end{eqnarray}
where $\gamma^a$ are constant matrices. In fact,  it produces  Klein-Gordon  equation if
\begin{eqnarray}\label{eq:Gam}
\{\gamma^a, \gamma^b\}  = 2 \eta^{a b} \, .
\end{eqnarray}
In expression  (\ref{eq:Gam}) we recognize relation  for $\gamma$-matrices.
Consequently, in that case  the  field  equations  consists of  Dirac like equation  plus supplementary condition
\begin{eqnarray}\label{eq:DllKGsc}
 (i \hat{\partial}  - m  )^\alpha{}_\beta  (P_i)^{\beta   \tilde{B} }{}_{\gamma  \tilde{C} } \Psi^{\gamma   \tilde{C} }  (x) = 0   \,  ,    \qquad
 ( P_j )^{\alpha  \tilde{A} }{}_{ \beta  \tilde{B} }    \Psi^{ \beta  \tilde{B} } (x)  =  0   \, ,   \qquad   (j \ne i  ) \, .
\end{eqnarray}
These   equations  can be combined  into  one equation  as linear combination of  Dirac  like  equation and supplementary conditions
\begin{eqnarray}\label{eq:coN1I}
\Big[  ( i {\hat{\partial}}  -   m  ) P_i  +  \sum_{j \ne i}   C_j P_j    \Big]^A{}_B  \Psi^B  (x)  =  0    \,  .  \qquad
\end{eqnarray}

Since  in general   ${\hat{\partial}} $  commute  or anticommute  with all projectors   in both Klein-Gordon and Dirac case
coefficients $C_i$  can depend on both   ${\hat{\partial}} $   and $m$    and we can take   $C_i =  c_i i {\hat{\partial}} +  \gamma_i m $, where $c_i$  and $\gamma_i$.   are numbers.

Note that all  operators    $P_i$  can contain singularities $\partial^{- 2}$.   If we  multiply   equations     (\ref{eq:coN1IKG})  and   (\ref{eq:coN1I})  with $\partial^2$  we can obtain   regular  equations   but with  higher derivatives.  In Refs.\cite{OS, N}   this problem was resolved  with  the  help of   root method.  It means that we can chose coefficients $C_i $
so that  (\ref{eq:coN1IKG})  and   (\ref{eq:coN1I}) become  second (or first) order  regular equations.
This procedure is not trivial  which can be seen  later in the example of Rarita-Schwinger case.

\section{Massive Dirac and vector   fields   }
\setcounter{equation}{0}

We will shortly repeat solutions for Dirac and vector  fields obtained in  Ref.\cite{BS1}, since Rarita-Schwinger field is combination of these two cases.
For the difference  of  Ref.\cite{BS1},  the case of the vector field  we will solved using projectors method. We will use  this solution  in the   Rarita-Schwinger case.

\subsection{Dirac  field   }

For   Dirac  field   $\Psi^A (x) \to \psi^\alpha (x)$ and  representation of  spin operator  is defined in   (\ref{eq:gLr21}).
Therefore,
\begin{eqnarray}\label{eq:gLr2f1}
( f_i )^\alpha{}_\beta   = \frac{1}{2} \, \varepsilon_{i j k} ( f_{j k} )^\alpha{}_\beta   = \frac{i}{4}  \varepsilon_{i j k} ( \gamma_j  \gamma_k )^\alpha{}_\beta   \, ,  \qquad
{\cal S}^\alpha{}_\beta  =   ( f_i^2 )^\alpha{}_\beta   =  \frac{3}{4 }  \delta^\alpha_\beta        \, .
\end{eqnarray}
The spin equation  (\ref{eq:BSms})  takes the form
\begin{eqnarray}\label{eq:BSs1d}
  {\cal S}^\alpha{}_\beta      \psi^\beta  (k) =  \lambda   \psi^\alpha  (k)   \, ,  \qquad   \lambda  \equiv  s (s + 1)
\end{eqnarray}
and since $ {\cal S}^\alpha{}_\beta$ is diagonal we obtain
\begin{eqnarray}\label{eq:}
\det (  {\cal S} - \lambda  )^\alpha{}_\beta
 =   \Big( \lambda -  \frac{3}{4}  \Big)^4 =     0             \,  .
\end{eqnarray}
Therefore,  $\lambda  =  \frac{3 }{4}$  which produces spin  $ s = \frac{1 }{2}$.  According to exponent  in last expression this is four dimensional representation,
which means that field $\psi^\alpha  (k) $  has four degrees of freedom.
There is only one trivial projection operator $P^\alpha{}_\beta  =   \delta^\alpha_\beta$.

We can go from standard momentum  $k^a$  to arbitrary one $p^a$  and then to coordinate representation.  So, we obtain standard
Klein-Gordon equation for all components $(\partial^2 + m^2  )  \psi^\alpha  (x) =0 $.
According to (\ref{eq:DllKG})   we can linearize  it  in the form
\begin{eqnarray}\label{eq:DiE}
(i \hat{\partial}   - m ) \psi^\alpha (x) = 0 \, ,  \qquad   (  \hat{\partial}  =  \gamma^a \partial_a ) \, .
\end{eqnarray}
Therefore, for fields with  spin $s = \frac{1 }{2}$  in expression  (\ref{eq:DiE}) we   recognize  Dirac equation.

\subsection{Vector  field   }

For vector fields we have   $\Psi^A  \to V^a $, and  representation of spin operator  (\ref{eq:gLr21})  is
\begin{eqnarray}\label{eq:mabV}
(s_{i j})^a{}_b = i \Big( \delta^a_i \, \eta_{j b}  - \delta^a_j \, \eta_{i b}  \Big) \, ,  \qquad
 (s_i)^a{}_b   =   \frac{1}{2} \, \varepsilon_{i j k} (S_{j k})^a{}_b  = i \, \varepsilon_{i j k}   \delta^a_j \, \eta_{k b}     \, ,
\end{eqnarray}
which produces
\begin{eqnarray}\label{eq:}
{\cal S}^a{}_b   =    (s^2_i)^a{}_b  =   (s_i)^a{}_c  (s_i)^c{}_b  =  - 2  \delta_{j k}     \delta^a_j \, \eta_{k b}    \, .
\end{eqnarray}

According to  (\ref{eq:IrrPa})  there are  two projectors
\begin{eqnarray}\label{eq:PiAB2}
( \pi_0 )^a{}_b (k)  =  \delta^a_b    -  \frac{  {\cal S}^a{}_b   }{   2   }   =   \delta^a_0  \delta^0_b         \, ,  \qquad
( \pi_1 )^a{}_b  (k)  =  \frac{  {\cal S}^a{}_b   }{   2   } =  \delta^a_b   -  \delta^a_0  \delta^0_b         \, .
\end{eqnarray}
In terms of projectors we have
\begin{eqnarray}\label{eq:SabS1}
 {\cal S}^a{}_b   =   2 \, (\pi_1)^a{}_b   \, ,   \qquad   \delta^a_b =  (\pi_0)^a{}_b  + (\pi_1)^a{}_b               \, ,
\end{eqnarray}
and consequently
\begin{eqnarray}\label{eq:SabS2}
 {\cal S}^a{}_b    - \lambda \delta^a_b   =  -  \lambda \Big(  \pi_0  \Big)^a{}_b   +  ( 2 - \lambda )  \Big(   \pi_1 \Big)^a{}_b    \, .
\end{eqnarray}
We will denote the subspace projected by $\pi_i$ as $\xi_i \, \,  (i = 0, 1)$. Then,  since $ dim \, \xi_0 = 1 $  and  $ dim \, \xi_1 = 3 $  we obtain
\begin{eqnarray}\label{eq:SabSd}
\det \Big(  {\cal S}^a{}_b   - \lambda \delta^a_b  \Big)
=  -  \lambda  ( 2 - \lambda )^3   \det \pi_0  \det \pi_1   =  0          \, ,
\end{eqnarray}
which produces $\lambda_0 = 0 $  and  $\lambda_1 = 2 $. Using  relation  $\lambda =  s (s + 1)$  we get  $s_0 = 0$ and $s_1 = 1$.

So, there are two  representations,   one  $V_0^a$  with spin-$0$   and the other  $V_1^a$  with spin-$1$  defined  as
\begin{eqnarray}\label{eq:}
V_0^a  (k)  =  (\pi_0 )^a{}_b   V^b    \, ,  \qquad    V_1^a  (k)  =  (\pi_1 )^a{}_b   V^b  \,  .
\end{eqnarray}
They satisfy   supplementary condition
\begin{eqnarray}\label{eq:}
(\pi_1 )^a{}_b   V_0^b  (k) =  0   \, ,  \qquad    (\pi_0 )^a{}_b   V_1^b  (k)   =   0    \,  .
\end{eqnarray}

Expressions for projectors in terms  of  rest frame momentum have a form
\begin{eqnarray}\label{eq:}
( \pi_0 )^a{}_b (k)   =  \frac{k^a k_b}{k^2}                 \, ,  \qquad
( \pi_1 )^a{}_b  (k)    =    \delta^a_b   -   \frac{k^a k_b}{k^2}   =   \delta^a_b   -    ( \pi_0 )^a{}_b (k)       \, .
\end{eqnarray}
In arbitrary  frame, with  $k_a \to p_a  \to i \partial_a$    they become longitudinal and   transversal projectors
\begin{eqnarray}\label{eq:}
( \pi_0 )^a{}_b  =  (P^L)^a{}_b       =  \frac{\partial^a \partial_b}{\partial^2}      \, ,   \qquad
( \pi_1 )^a{}_b    =  (P^T)^a{}_b     =  \delta^a_b -    \frac{\partial^a \partial_b}{\partial^2}        \, .
\end{eqnarray}

The  supplementary  conditions  are automatically satisfied  since   $\pi_0$   and  $\pi_1$   are projectors
\begin{eqnarray}\label{eq:scon01}
(\pi_1 )^a{}_b   V_0^b  (x) =  0   \, ,  \qquad    (\pi_0 )^a{}_b   V_1^b  (x)   =   0    \,  .
\end{eqnarray}
In arbitrary frame we should include the Klein-Gordon equations for both components
\begin{eqnarray}\label{eq:KG01}
( \partial^2  +  m^2  )  V_0^a (x)  = 0   \,  ,   \qquad    ( \partial^2  +  m^2  )  V_1^a (x)  = 0            \, .
\end{eqnarray}
So, solutions for spin-$0$  and spin-$1$ fields $V_0^a$  and  $V_1^a$  satisfy relations (\ref{eq:scon01})  and  (\ref{eq:KG01}).

We are looking for an equations whose solutions are  $V_0^a$  and  $V_1^a$.
Therefore, complete field equation are:  for  spin-$0$  field  $v^a$
\begin{eqnarray}\label{eq:KG01p}
( \partial^2  +  m^2  )  (\pi_0 )^a{}_b   v^b  (x)   = 0   \,  ,   \qquad    (\pi_1 )^a{}_b   v^b  (x) =  0          \, ,
\end{eqnarray}
and for  spin-$1$  field  $A^a$
\begin{eqnarray}\label{eq:KG011}
   ( \partial^2  +  m^2  )   (\pi_1 )^a{}_b   A^b  (x)     = 0    \,  ,   \qquad         (\pi_0 )^a{}_b   A^b  (x)   =   0      \, .
\end{eqnarray}
Consequently,  for both spins field  equations  consists of Klein-Gordon  equation for the vector field plus supplementary condition.

These   field   equations  can be combined  into  one equation  as linear combination of  Klein-Gordon equation and supplementary conditions  for  spin-$0$  and  spin-$1$   fields
\begin{eqnarray}\label{eq:coN12v}
\Big[  (  \partial^2  +  m^2  ) \pi_0  +   C  \pi_1       \Big]^a{}_b  v^b  (x)  =  0     \,  ,   \qquad
\Big[  (  \partial^2  +  m^2  )   \pi_1    +  D  \pi_0   \Big]^a{}_b   A^b  (x)  =  0    \,  .  \qquad
\end{eqnarray}

\subsubsection{Regular form  of vector equations }

As explained in general case   we can chose coefficients $C $  and $D$ so that vector  equations   become   regular.
With the help of  relation    $C = a \partial^2 + \alpha  m^2$   we can separate kinetic and mass terms of  equations   (\ref{eq:coN12v})  for spin-$0$ field  in the form
\begin{eqnarray}\label{eq:}
 ( K_0)^a{}_b  =    \partial^2 \Big(  \pi_0  +   a  \pi_1       \Big)^a{}_b       \,  ,  \qquad
( M_0 )^a{}_b  =   m^2   \Big(   \pi_0  +   \alpha  \pi_1       \Big)^a{}_b      \,  .  \qquad
\end{eqnarray}
and  with the help of     $D = b \partial^2 + \beta  m^2$  for spin-$1$ field  in the form
\begin{eqnarray}\label{eq:}
 ( K_1)^a{}_b  =  \partial^2 \Big(  \pi_1  +   b   \pi_0       \Big)^a{}_b       \,  ,  \qquad
( M_1 )^a{}_b  =   m^2    \Big(  \pi_1  +   \beta  \pi_0       \Big)^a{}_b      \,  .  \qquad
\end{eqnarray}

Thanks to the presence  of operator  $\partial^2$  kinetic terms are regular.
Using $\pi_1 = 1 - \pi_0$  we can kill singular parts $(\pi_0)^a{}_b$ in  mass terms   with conditions
\begin{eqnarray}\label{eq:}
 1 - \alpha = 0 \, ,  \qquad   -  1 + \beta = 0 \, ,  \qquad
\end{eqnarray}
so that   the  regular  field    equations  take a form
\begin{eqnarray}\label{eq:GFvf}
\Big[    \Big( a  + (1 - a)  \pi_0 \Big)  \partial^2  +  m^2  \Big]^a{}_b  v^b  = 0    \, ,   \qquad
\Big[    \Big( 1  + (b - 1 )  \pi_0 \Big)  \partial^2  +  m^2  \Big]^a{}_b  A^b  = 0                 \,  .
\end{eqnarray}

These are the most general vector field equations for spin-$0$  and  spin-$1$  fields with arbitrary coefficients $a$  and $b$.
Multiplying first equation with $\pi_0$   and second one with $\pi_1$  we obtain coefficients independent relations
\begin{eqnarray}\label{eq:}
\Big(  \partial^2  +  m^2  \Big)    (\pi_0)^a{}_b  v^b  = 0    \, ,   \qquad
  \Big( \partial^2  +  m^2  \Big)     (\pi_1)^a{}_b  A^b   = 0                 \,  .
\end{eqnarray}

Introducing new variables
\begin{eqnarray}\label{eq:}
\varphi =  \partial_a  (\pi_0)^a{}_b  v^b  \,  ,  \qquad    U^a =  (\pi_1)^a{}_b  A^b  \, ,
\end{eqnarray}
we obtain standard scalar and vector  equation
\begin{eqnarray}\label{eq:KGpU}
\Big(  \partial^2  +  m^2  \Big)   \varphi = 0    \, ,   \qquad
  \Big( \partial^2  +  m^2  \Big)    U^a   = 0                 \,  ,
\end{eqnarray}
 where the vector field $U^a$ is constrained $\partial_a U^a  = 0$.

We can consider Eqs.(\ref{eq:GFvf})   as gauge fixed version of standard equations (\ref{eq:KGpU}).

\section{Massive   Rarita-Schwinger  field  }
\setcounter{equation}{0}

For Rarita-Schwinger   fields we have   $A, B \to (a \alpha), (b \beta)$ and $\Psi^A   \to \psi^{a \alpha} $  where $a, b$ are vector and  $\alpha, \beta$  spinor indices, so that spin equation takes the form
\begin{eqnarray}\label{eq:COPfv}
 {\cal S}^{a \alpha}{}_{b \beta}    \psi^{b \beta}    = \lambda  \psi^{a \alpha}       \, ,             \qquad
  {\cal S}^{a \alpha}{}_{b \beta}  =    (S^2_i)^{a \alpha}{}_{b \beta}   \, ,   \qquad    \lambda  =   s( s + 1)  \, .
\end{eqnarray}

\subsection{Spin operator for  Rarita- Schwinger  field   }

According  to   (\ref{eq:BEder})  spin operator for Rarita- Schwinger  field
\begin{eqnarray}\label{eq:SRTRS0}
 ( S_{a b}  )^{c \alpha}{}_{d \beta}
=   ( s_{a b}  )^{c }{}_{d}  \delta^\alpha_\beta  +    \delta^c_d    ( f_{a b}   )^{\alpha}{}_{ \beta}  \,  ,
\end{eqnarray}
can be express in terms of  spin operators for  Dirac fields   $( f_{a b}  )^{\alpha}{}_{ \beta} $  and spin operators  for  vector  fields  $ ( s_{a b}  )^{c }{}_{d} $,   introduced in  (\ref{eq:gLr21}).  Then we have
\begin{eqnarray}\label{eq:}
(S_i)^{c \alpha}{}_{d \beta}
=  \frac{1}{2} \, \varepsilon_{i j k} (S_{j k})^{c \alpha}{}_{d \beta}
=  ( s_{i}  )^{c }{}_{d}  \delta^\alpha_\beta  +    \delta^c_d    ( f_{i}   )^{\alpha}{}_{ \beta}    \, ,  \qquad
\end{eqnarray}
where for vector and Dirac fields
\begin{eqnarray}\label{eq:}
 (  s_{i}  )^{c }{}_{d}   =    \frac{1}{2}  \, \varepsilon_{i j k}  ( s_{j k}  )^{c }{}_{d}
 =  i \, \varepsilon_{i j k}  \delta^c_j \, \eta_{k d}   \,  ,  \qquad
 (  f_{i}   )^{\alpha}{}_{ \beta} =  \frac{1}{2}  \, \varepsilon_{i j k}   ( f_{j k}  )^{\alpha}{}_{ \beta}
 =   \frac{i}{4}   \, \varepsilon_{i j k}    ( \gamma_j \gamma_k )^\alpha{}_\beta   \, . \qquad
\end{eqnarray}
Consequently
\begin{eqnarray}\label{eq:}
(S^2_i)^{c \alpha}{}_{d \beta}   =   (S_i)^{c \alpha}{}_{e \gamma}   (S_i)^{e \gamma}{}_{d \beta}
=   ( s_i^2  )^{c }{}_d   \delta^\alpha_\beta +    \delta^c_d    ( f_i^2 )^{\alpha}{}_\beta
+  2 ( s_{i}  )^{c }{}_d    ( f_{i} )^{\alpha}{}_\beta        \, .
\end{eqnarray}

From   the cases of vector and Dirac fields  we had
\begin{eqnarray}\label{eq:}
(s^2_i)^a{}_b  =    - 2    \delta^a_i \, \eta_{i b}    \,  ,  \qquad
( f_i^2 )^\alpha{}_\beta   =  \frac{3}{4 }  \delta^\alpha_\beta        \, .
\end{eqnarray}
The mixed term has a form
\begin{eqnarray}\label{eq:}
 ( s_{i}  )^{c }{}_d    ( f_{i} )^{\alpha}{}_\beta
=   -    \frac{1}{4}   \delta^c_j \, \eta_{k d}  [ \gamma_j, \gamma_k ]^\alpha{}_\beta  \,  ,
\end{eqnarray}
so that
\begin{eqnarray}\label{eq:}
 {\cal S}^{a \alpha}{}_{b \beta}  =  (S^2_i)^{a \alpha}{}_{b \beta}
=   \Big(  \frac{3}{4}  \delta^a_b  - 2 \delta^a_i \, \eta_{i b} \Big)  \delta^\alpha_\beta
   -  \frac{1}{2}  \delta^a_i \, \eta_{j b} [ \gamma_i, \gamma_j ]^\alpha{}_\beta  \, .   \qquad
\end{eqnarray}
Using  relation  $  [ \gamma_j, \gamma_k ]  = 2  (\delta_{j k}  +   \gamma_j \gamma_k  ) $  we obtain Rarita-Schwinger  spin operator
\begin{eqnarray}\label{eq:sopRS}
 {\cal S}^{a \alpha}{}_{b \beta}
=  3 \Big(  \frac{1}{4}  \delta^a_b  -  \delta^a_i \, \eta_{i b} \Big)  \delta^\alpha_\beta
   -    \delta^a_i \, \eta_{j b} ( \gamma_i \gamma_j )^\alpha{}_\beta      \, .   \qquad
\end{eqnarray}

\subsubsection{Spin operator in terms of elementary projectors   }

To solve consistency condition we need to calculate  determinant of  $16 \times 16$ matrix   ${\cal S}^{a \alpha}{}_{b \beta} - \lambda  \delta^a_b \delta^\alpha_\beta$.  In order to simplify
calculations we will introduce projectors method.

We can rewrite equation   (\ref{eq:sopRS}) in the form
\begin{eqnarray}\label{eq:}
 {\cal S}^{a \alpha}{}_{b \beta}
=  \frac{3}{4}   (\pi_0)^a{}_b  \delta^\alpha_\beta    + \frac{15}{4}   (\pi_1)^a{}_b   \delta^\alpha_\beta     - 3  ( \pi)^{a \alpha}{}_{b \beta}      \, ,   \qquad
\end{eqnarray}
where  projectors of vector case  $(\pi_0)^a{}_b $   and  $(\pi_1)^a{}_b $  has been defined in   (\ref{eq:PiAB2})  and
\begin{eqnarray}\label{eq:PRpi}
  (\pi)^{a \alpha}{}_{b \beta}   =  \frac{1}{3}    \delta^a_i \, \eta_{j b} ( \gamma_i \gamma_j )^\alpha{}_\beta       \, .  \quad
\end{eqnarray}

Note that  $(\pi_0)^a{}_b $,  $(\pi_1)^a{}_b $  and  $\pi^a{}_b $  are  projectors since
\begin{eqnarray}\label{eq:}
 \pi_0^2 =   \pi_0        \, ,  \qquad   \pi_1^2 =   \pi_1        \, ,  \qquad    \pi^2 =   \pi        \, ,  \qquad
\end{eqnarray}
but we also have relations
\begin{eqnarray}\label{eq:}
 \pi_0 \pi_1  =  0         \, ,  \qquad   \pi_0 \pi = 0        \, ,  \qquad   \pi_1 \pi =   \pi        \, .   \qquad
\end{eqnarray}
So,  space projected by  $\pi$ is  subspace of  that   projected by  $\pi_1$,   $\xi \subset  \xi_1$.

We can construct three projectors   which have no intersection
\begin{eqnarray}\label{eq:}
P_{0 1} = \pi_0        \, ,  \qquad   P_{0 2}  =  \pi        \, ,  \qquad    P_1 =   \pi_1 - \pi        \, ,  \qquad
\end{eqnarray}
such that
\begin{eqnarray}\label{eq:}
P_n^2  = P_n          \, ,  \qquad   P_n P_m    =  \delta_{n m} P_n          \, ,  \qquad    n, m = \{ (0 1), (0 2), 1   \}        \, .  \qquad
\end{eqnarray}

The projector $\pi$  has a form
\begin{eqnarray}\label{eq:}
\pi^{a }{}_{b }  =
 \frac{1}{3}  \left| {\begin{array}{cccc}
 0    & 0    &  0   &  0   \\
 0   &   1   & - \gamma^1 \gamma_2   & - \gamma^1 \gamma_3   \\
 0   & - \gamma^2 \gamma_1   &   1   & - \gamma^2 \gamma_3  \\
  0  &  - \gamma^3 \gamma_1   & - \gamma^3 \gamma_2   &   1  \\
\end{array} } \right|  \,  ,
\end{eqnarray}
Because determinant of  each  $2 \times 2$  submatrix  is zero, the rank of $\pi$ is $1 \cdot 4 = 4$  where $1$ is rank of vector part and $4$  is rank of spinor part.

We will denote the subspace projected by  $\pi_0, \pi, \pi_1 $  with  $\xi_0, \xi, \xi_1$  and by   $P_n$ with  $\Xi_n$.   Consequently  since   $\xi \subset  \xi_1$ and
$dim \, \xi_0 = 4$, $dim \, \xi = 4$  and   $dim \, \xi_1 = 12$  we   have     $dim \, \Xi_{0 1} = dim  \, \Xi_{0 2} = 4$  and   $ dim  \,\Xi_1 = 8$.
As we will see,  field separated by  projector $P_1$  is spin-$\frac{3}{2}$  field with $8$ degrees of freedom.  There are two   spin-$\frac{1}{2}$  fields separated by projectors
$P_{0 1} = \pi_0$  and  $P_{0 2} = \pi$, each of them with $4$ degrees of freedom.

\subsubsection{Auxiliary  projectors   }

We can rewrite operators $\pi_0$   and $\pi$ in the form
\begin{eqnarray}\label{eq:p0PRpi}
(\pi_0)^{a \alpha}{}_{b \beta}  =   (\rho_0 )^{a \alpha}{}_\gamma      ( \rho_0)^\gamma{}_{b \beta}     \, ,   \qquad
  (\pi)^{a \alpha}{}_{b \beta}   =   \frac{1}{3}   (\rho )^{a \alpha}{}_\gamma      (\rho )^{b \gamma}{}_\beta    \, ,  \quad
\end{eqnarray}
where
\begin{eqnarray}\label{eq:rho0}
 (\rho_0 )^{a \alpha}{}_\gamma  =  \delta^a_0  ( \gamma^0 )^\alpha{}_\gamma  \, ,  \qquad    (\rho )^{a \alpha}{}_\gamma  =  \delta^a_i  (   \gamma^i  )^\alpha{}_\gamma   \,  ,   \qquad
 (\rho_0 + \rho)^{a \alpha}{}_\gamma  =     ( \gamma^a )^\alpha{}_\gamma    \, .
\end{eqnarray}

It is useful to introduce new  auxiliary  projectors as   mixed expressions
\begin{eqnarray}\label{eq:po0ast}
(\pi_{0 \ast})^a{}_b   =  \frac{1}{\sqrt{3}} \rho_0^a  ( \rho)_b      \, ,  \qquad
  (\pi_{\ast 0})^a{}_b     =    \frac{1}{\sqrt{3}}   \rho^a       ( \rho_0)_b     \, .  \quad
\end{eqnarray}

Then
\begin{eqnarray}\label{eq:s}
(  \pi_0 + \sqrt{3} \pi_{0 \ast})^a{}_b   =   \rho_0^a  \gamma_b      \, ,  \qquad
  (\pi +  \sqrt{3}  \pi_{\ast 0})^a{}_b     =       \rho^a       \gamma_b     \, ,  \quad   \nonumber \\
(  \pi + \sqrt{3} \pi_{0 \ast})^a{}_b   =   \gamma^a  \rho_b      \, ,  \qquad
  (\pi_0  +  \sqrt{3}  \pi_{\ast 0})^a{}_b     =       \gamma^a     (\rho_0)_b     \, ,  \quad
\end{eqnarray}
so that
\begin{eqnarray}\label{eq:sf}
\Big(  \pi_0 + \pi +  \sqrt{3} (  \pi_{0 \ast}  +  \pi_{\ast 0} )  \Big)^a{}_b   =   \gamma^a  \gamma_b      \, .  \quad
\end{eqnarray}

Using relations
\begin{eqnarray}\label{eq:}
   \rho_0^a  ( \rho_0)_a = 1      \, ,  \qquad       \rho^a       \rho_b   =  3   \, ,  \quad     \rho_0^a   \rho_a =  0      \, ,  \qquad
\end{eqnarray}
we obtain
\begin{eqnarray}\label{eq:}
\pi_{0 \ast}^2     =  0      \, ,  \qquad   \pi_{\ast 0}^2    =   0   \, , \quad
\pi_{0 \ast}  \pi_{\ast 0}      = \pi_0      \, ,  \qquad   \pi_{\ast 0} \pi_{0 \ast}   =   \pi   \, , \quad     \nonumber \\
\pi_{0 \ast}  \, \pi      = \pi_{0 \ast}       \, ,  \qquad   \pi_{\ast 0} \, \pi_0      =   \pi_{\ast 0}  \, . \quad
\end{eqnarray}

We also have
\begin{eqnarray}\label{eq:}
  ( P_1)^a{}_b  ( \rho_0)^b = 0      \, ,  \qquad      ( P_1)^a{}_b     \rho_b   =  0     \, ,  \qquad
\end{eqnarray}
so that
\begin{eqnarray}\label{eq:}
   P_1  \pi_{0 \ast}   = 0      \, ,  \qquad      P_1    \pi_{\ast 0}    =  0      \, .  \qquad
\end{eqnarray}

The  expressions  $P_1, \pi_0, \pi, \pi_{0 \ast},  \pi_{\ast 0}$ are just  projectors   introduced in Ref.\cite{N}, but here in frame of standard momentum.
Later   we will obtain  corresponding   expressions in arbitrary frame and  show that they coincident to that of Ref.\cite{N}.

\subsection{Rarita-Schwinger  equation  in frame of standard momentum }

Let us first derive Rarita-Schwinger  equation  in frame of standard momentum  and then we will boost it.

\subsubsection{Rarita-Schwinger  projectors in terms of elementary and  auxiliary projectors   }

With the help of  expression    $\delta^a_b   =   ( \pi_0 )^a{}_b  + ( \pi_1 )^a{}_b    $   we  obtain
\begin{eqnarray}\label{eq:}
  {\cal S}^{a }{}_{b }   -  \lambda   \delta^a_b
 =  \Big( \frac{3}{4} - \lambda \Big)  (\pi_0)^a{}_b     + \Big( \frac{15}{4} -  \lambda \Big)  (\pi_1)^a{}_b   -  3   (\pi)^a{}_b        \,  .  \qquad
\end{eqnarray}
Then,  using  relation
\begin{eqnarray}\label{eq:}
 \Big( \frac{3}{4} - \lambda \Big)  - \Big( \frac{15}{4} -  \lambda \Big)       =  -  3   \,  ,  \qquad
\end{eqnarray}
we   can rewrite above equation  as
\begin{eqnarray}\label{eq:}
  {\cal S}^{a }{}_{b }   -  \lambda   \delta^a_b
 =   \Big( \frac{3}{4} - \lambda \Big) ( P_0 )^a{}_b    + \Big( \frac{15}{4} -  \lambda \Big) ( P_1 )^a{}_b  \,  ,  \qquad
\end{eqnarray}
where  we introduced  expressions
\begin{eqnarray}\label{eq:}
 ( P_0 )^a{}_b   =    (\pi_0 + \pi)^a{}_b      \, ,  \quad
  ( P_1 )^a{}_b =    (\pi_1  -  \pi)^a{}_b       \,  .  \qquad
\end{eqnarray}
It is easy to check that  $P_0$  and $P_1$ are projectors
\begin{eqnarray}\label{eq:}
  P_0^2  =    P_0      \, ,  \quad
 P_1^2   =     P_1     \, ,  \quad
  P_0 P_1   =     0     \, ,  \quad  P_0 + P_1 = 1 \, .
\end{eqnarray}

\subsubsection{Spectrum  of vector-spinor fields  and corresponding  supplementary  conditions for vector-spinor field   }

Therefore, the  spin equation (\ref{eq:COPfv})  in the rest frame  takes the  form
\begin{eqnarray}\label{eq:}
\Big[  \Big( \frac{3}{4} - \lambda \Big) P_0     +  \Big( \frac{15}{4} -  \lambda \Big)  P_1 \Big]  \psi  = 0    \,  .   \qquad   \lambda   \equiv s (s + 1)
\end{eqnarray}
The   consistency condition   produces
\begin{eqnarray}\label{eq:}
\det (  {\cal S} - \lambda  )^{a \alpha}{}_{b \beta}  =   \Big( \frac{3}{4} - \lambda \Big)^{d_0}   \Big( \frac{15}{4} -  \lambda \Big)^{d_1}  \det  P_0 \, \det P_1  = 0      \,  ,
\end{eqnarray}
where $d_0  = d_1  =8$ are dimensions of subspaces defined by $P_0$   and  $P_1$.
The solutions for eigenvalues  are  $\lambda_0    =    \frac{3}{4}  \, ,  \,     \lambda_1    = \frac{15}{4}   $     and for spins    $ s_0 = \frac{1}{2}  \, , \,   s_1 =  \frac{3}{2} $.

The fields  corresponding to one  spin-$\frac{3}{2} $   and  two spin-$\frac{1}{2}$   fields  are defined by projection operators
\begin{eqnarray}\label{eq:}
 \psi_1 =   P_1  \psi =    (\pi_1 - \pi )  \psi  \, ,  \qquad   \psi_{0 1} =   P_{0 1} \psi  = \pi_0 \psi \, ,    \qquad   \psi_{0 2} =   P_{0 2} \psi  = \pi \psi     \, .  \quad
\end{eqnarray}
Consequently, in the frame of standard momentum   supplementary  conditions are

1.  for field   $\psi_1$     with   $s = \frac{3}{2}$
\begin{eqnarray}\label{eq:Suc32}
 \pi_0   \psi_1  (k) =  0     \,  ,  \qquad    \pi  \psi_1  (k) =  0    \,  ,  \qquad
\end{eqnarray}

2. for field   $\psi_{0 1}$     with   $s = \frac{1}{2}$
\begin{eqnarray}\label{eq:Suc121}
 \pi  \psi_{0 1}  (k)  =  0    \, ,  \qquad  P_1   \psi_{0 1}   (k)  =  0      \, ,  \qquad
\end{eqnarray}

3.   for field   $\psi_{0 2}$     with   $s = \frac{1}{2}$
\begin{eqnarray}\label{eq:Suc122}
 \pi_0    \psi_{0 2}  (k)  =  0   \, ,  \qquad       P_1  \psi_{0 2}   (k)  =  0    \, ,  \qquad
\end{eqnarray}

\section{Rarita-Schwiger equation in   arbitrary  frame   }
\setcounter{equation}{0}

In order to obtain irreducible representations   in arbitrary frame we should boost corresponding equation of the rest frame.
It will help us to  obtain Rarita-Schwiger equation in   arbitrary  frame.

\subsection{From standard momentum frame  to arbitrary  frame   }

We will use expressions for vector boosts components
\begin{eqnarray}\label{eq:}
 {}_1  {\cal B}^a{}_0   (\mathbf{p})   =  \frac{p^a}{m}    \, ,   \qquad   {}_1   ({\cal B}^{-1})^0{}_a   (\mathbf{p})   =    \frac{p_a}{m}    \, ,
\end{eqnarray}
and for spinor boosts components
\begin{eqnarray}\label{eq:}
{}_{\frac{1}{2} }  {\cal F}^\alpha{}_\beta  (\mathbf{p})
= \frac{1}{\sqrt{ 2 m (\omega_{\mathbf{p}}  + m )} }\,  \Big( m     + {\hat p}   \gamma^0   \,  \Big)^\alpha{}_\beta   \, ,   \qquad
\end{eqnarray}
where
\begin{eqnarray}\label{eq:}
  p_a = \{\omega_{ \mathbf{p}}, p_i \} \, ,  \qquad \omega_{ \mathbf{p}} =  \sqrt{\mathbf{p}^2 + m^2}        \, .  \qquad
\end{eqnarray}

It is useful  to derive first expressions for auxiliary components    (\ref{eq:rho0})
\begin{eqnarray}\label{eq:}
  (\rho_0 )^{a \alpha}{}_\beta   (p)
=  {}_1 {\cal B}^a{}_b  (\mathbf{p}) \,  {}_{\frac{1}{2} }    {\cal F}^\alpha{}_\gamma (\mathbf{p})   (\rho_0 )^{b \gamma}{}_\delta   (k)
 \, ( {}_{\frac{1}{2} }  {\cal F}^{-1})^\delta{}_\beta  (\mathbf{p})    \, .
\end{eqnarray}
 Using relation
\begin{eqnarray}\label{eq:CBm1CB2n}
\Big( {}_{\frac{1}{2}}  {\cal F} ( \mathbf{p}) \gamma^a  {}_{\frac{1}{2}}  {\cal F}^{-1} ( \mathbf{p} )  \Big)^\alpha{}_\beta =    {}_1 ({\cal B}^{-1})^a{}_b ( \mathbf{p} ) ( \gamma^b )^\alpha{}_\beta  \, ,
\end{eqnarray}
we obtain
\begin{eqnarray}\label{eq:}
  (\rho_0 )^{a \alpha}{}_\beta   (p)     =   (P_L)^a{}_b   ( \gamma^b )^\alpha{}_\beta     \, ,
\end{eqnarray}
and    with the help of the relation  $  (\rho )^{a \alpha}{}_\gamma (k) =  (   \gamma^a  )^\alpha{}_\gamma  -     (\rho_0 )^{a \alpha}{}_\gamma $ we have
\begin{eqnarray}\label{eq:}
  (\rho )^{a \alpha}{}_\gamma  (p)   =   (P_T)^a{}_b   ( \gamma^b )^\alpha{}_\beta         \, .    \qquad
\end{eqnarray}
Note that
\begin{eqnarray}\label{eq:hprh}
\hat{p} \, \rho_0^a =  \rho_0^a \, \hat{p} =  p^a \, ,  \qquad     \hat{p} \, \rho^a =  - \rho^a \, \hat{p}    \, .  \qquad
\end{eqnarray}

Now, using   (\ref{eq:p0PRpi})   and   (\ref{eq:po0ast}) we obtain
\begin{eqnarray}\label{eq:pi0pip}
(\pi_0)^a{}_b  (p) =   \rho_0^a  ( \rho_0)_b    =  (P_L)^a{}_c    (P_L)_{b d}   (\gamma^c  \gamma^d )^\alpha{}_\beta    =    \frac{\partial^a  \partial_b}{\partial^2}  =   (P_L)^a{}_b      \, ,       \nonumber \\
(\pi_1)^a{}_b  (p)   =  \delta^a_b -  (P_L)^a{}_b     =   (P_T)^a{}_b    \, ,   \nonumber \\
 (\pi)^a{}_b   (p)  =    \frac{1}{3}   \rho^a   \rho_b   =    \frac{1}{3}   (P_T)^a{}_c  (P_T)^b{}_d ( \gamma^c \gamma_d )^\alpha{}_\beta                 \, ,  \quad
\end{eqnarray}
and
\begin{eqnarray}\label{eq:po0astp}
(\pi_{0 \ast})^a{}_b (p)   =  \frac{1}{\sqrt{3}} \rho_0^a  ( \rho)_b    =  \frac{1}{\sqrt{3}}    (P_L)^a{}_c  (P_T)^b{}_d ( \gamma^c \gamma_d )^\alpha{}_\beta                \, ,  \qquad            \nonumber \\
  (\pi_{\ast 0})^a{}_b    (p)  =    \frac{1}{\sqrt{3}}   \rho^a    ( \rho_0)_b   =  \frac{1}{\sqrt{3}}   (P_T)^a{}_c  (P_L)^b{}_d ( \gamma^c \gamma_d )^\alpha{}_\beta        \, .  \quad
\end{eqnarray}

Using  (\ref{eq:hprh}) we   have
\begin{eqnarray}\label{eq:hprhp}
\hat{p} \, (\pi_{0 \ast})^a{}_b =  -  (\pi_{0 \ast})^a{}_b \, \hat{p}     \, ,  \qquad     \hat{p} \,  (\pi_{\ast 0})^a{}_b =  -  (\pi_{\ast 0})^a{}_b \, \hat{p}    \, .  \qquad
\end{eqnarray}

Finally, Rarita-Schwinger projectors in arbitrary frame are
\begin{eqnarray}\label{eq:}
(P_0 )^{a \alpha }{}_{b \beta}  (p)  =   (\pi_0)^a{}_b    \delta^\alpha_\beta     +  \frac{1}{3} ( \pi_1 )^a{}_c   ( \pi_1 )_b{}^d    ( \gamma^c \gamma_d )^\alpha{}_\beta     \,  ,   \qquad      \nonumber \\
(P_1 )^{a \alpha }{}_{b \beta}   (p)  =   (\pi_1)^a{}_b    \delta^\alpha_\beta   -  \frac{1}{3} ( \pi_1 )^a{}_c   ( \pi_1 )_b{}^d    ( \gamma^c \gamma_d )^\alpha{}_\beta      \,  .  \qquad
\end{eqnarray}

\subsection{Spin-$\frac{1}{2}$ Rarita-Schwinger  equations in arbitrary  frame  }

Let us  consider the case of spin-$\frac{1}{2}$  fields   $\psi_{0 1}$  and  $\psi_{0 1}$.   In frame of standard momentum   supplementary  conditions   on fields   $\psi_{0 1} (k)$  and    $\psi_{0 2} (k)$
have a form  (\ref{eq:Suc121})  and  (\ref{eq:Suc122}).
We can  turn to   arbitrary  frame  in standard way   $k_a \to p_a  \to i \partial_a$.  Then fields  $\psi_{0 1} (x) = \pi_0 \psi (x)$   and  $\psi_{0 2} (x)  = \pi \psi (x)$
must satisfy the same  supplementary  conditions  as in  the   frame of standard momentum
\begin{eqnarray}\label{eq:sc01}
 \pi ( \psi_{0 1} )^{a \alpha} (x)  =  0    \, ,  \qquad  P_1  ( \psi_{0 1} )^{a \alpha}  (x)  =  0      \, ,  \qquad    \nonumber \\
 \pi_0   ( \psi_{0 2} )^{a \alpha}  (x)  =  0   \, ,  \qquad       P_1 ( \psi_{0 2} )^{a \alpha}  (x)  =  0    \, ,  \qquad
\end{eqnarray}
and  according to (\ref{eq:DllKG})   Dirac like equations
\begin{eqnarray}\label{eq:RSD1}
( i {\hat{\partial}}  -   m  )  \psi_{0 1}^a (x)  = 0   \,  ,   \qquad    ( i {\hat{\partial}}  -   m  )  \psi_{0 2}^a (x)  = 0               \, .
\end{eqnarray}
Consequently,   we  can introduce   two sets of  spin-$\frac{1}{2}$ Rarita-Schwinger  equations in arbitrary  frame
\begin{eqnarray}\label{eq:con11c}
( i {\hat{\partial}}  -   m  ) (\pi_0)^a{}_b  \psi^b (x)  = 0  \, ,   \qquad   (P_1)^a{}_b  \psi^{b \alpha} (x) =  0     \,  ,  \qquad  (\pi )^a{}_b \psi^{b \alpha} (x) =  0     \,  ,   \qquad    \nonumber \\
( i {\hat{\partial}}  -   m  ) (\pi)^a{}_b  \psi^b (x)  = 0  \, ,   \qquad   (P_1)^a{}_b  \psi^{b \alpha} (x) =  0     \,  ,  \qquad  (\pi_0 )^a{}_b \psi^{b \alpha} (x) =  0     \,  ,  \qquad
\end{eqnarray}
so  that by definition  $\psi_{0 1}^a (x)$  and  $\psi_{0 2}^a (x)$  are their solutions.
It show that field  equations  consists of  Dirac equation for the vector-spinor field plus supplementary condition.

Every  set   of  spin-$\frac{1}{2}$  Rarita-Schwinger  equations  can be combined  into  one equation  as linear combination of  Dirac equation and supplementary conditions
\begin{eqnarray}\label{eq:coN120}
\Big[  ( i {\hat{\partial}}  -   m  ) \pi_0  +   A  P_1    +  B  \pi   \Big]^a{}_b  \psi^{b \alpha} (x)  =  0     \,  ,   \qquad              (I)    \nonumber \\
\Big[  ( i {\hat{\partial}}  -   m  )   \pi  +   C  P_1    +  D  \pi_0   \Big]^a{}_b  \psi^{b \alpha} (x)  =  0    \,  .  \qquad            (II)
\end{eqnarray}

\subsection{Spin-$\frac{3}{2}$  Rarita-Schwinger  equation in arbitrary  frame  }

Let us  consider the case of spin-$\frac{3}{2}$  field $\psi_1$.   In frame of standard momentum   supplementary  conditions   on field   $\psi_1  (k)$  have a form  (\ref{eq:Suc32}).
When we turn to arbitrary  frame using  $k_a \to p_a  \to i \partial_a$  then    field  $\psi_1 (x)$  must satisfy the same  supplementary conditions
\begin{eqnarray}\label{eq:con11}
 (\pi_0 )^a{}_b  \psi_1^{b \alpha} (x) =  0     \,  ,  \qquad   ( \pi )^a{}_b \psi_1^{b \alpha} (x) =  0    \,  .  \qquad
\end{eqnarray}
These constraints  eliminate the unwanted    spin-$\frac{1}{2}$ components of the vector-spinor  field  $\psi^{a \alpha} $.

With the help of   (\ref{eq:p0PRpi})   we can rewrite them   in the form
\begin{eqnarray}\label{eq:}
 (\rho_0 )_b  \psi_1^{b \alpha} (x) =  0     \,  ,  \qquad   \rho_b \psi_1^{b \alpha} (x) =  0    \,  ,  \qquad
\end{eqnarray}
and using   (\ref{eq:po0ast})  as
\begin{eqnarray}\label{eq:piastn}
 (\pi_{\ast 0} )^a{}_b  \psi_1^{b \alpha} (x) =  0     \,  ,  \qquad   ( \pi_{0 \ast} )^a{}_b \psi_1^{b \alpha} (x) =  0    \,  .  \qquad
\end{eqnarray}
On-shell,  these  supplementary  conditions   have a form
\begin{eqnarray}\label{eq:con11o}
 \partial_b  \psi_1^{b \alpha} (x) =  0     \,  ,  \qquad   \gamma_b \psi_1^{b \alpha} (x) =  0    \,  .  \qquad
\end{eqnarray}

According to (\ref{eq:DllKG})  the Dirac like equation is
\begin{eqnarray}\label{eq:RSD1}
( i {\hat{\partial}}  -   m  )  \psi_1^a (x)  = 0  \, .
\end{eqnarray}

Consequently,    since $\psi_1  = P_1 \psi$  from   (\ref{eq:con11}),   (\ref{eq:piastn})  and  (\ref{eq:RSD1})    we obtain  spin-$\frac{3}{2}$ Rarita-Schwinger  equations in arbitrary  frame
\begin{eqnarray}\label{eq:con11c}
( i {\hat{\partial}}  -   m  ) (P_1)^a{}_b  \psi^b (x)  = 0  \, ,   \qquad   ( \pi_0  )^a{}_b  \psi^{b \alpha} (x) =  0     \,  ,  \qquad  (\pi )^a{}_b \psi^{b \alpha} (x) =  0   \, ,  \nonumber \\
 (\pi_{\ast 0} )^a{}_b  \psi^{b \alpha} (x) =  0     \,  ,  \qquad   ( \pi_{0 \ast} )^a{}_b \psi^{b \alpha} (x) =  0   \,  .  \qquad
\end{eqnarray}
such  that by definition  $\psi_1^a (x)$  is  their solutions.
It show that field  equations  consists of  Dirac equation for the vector-spinor field plus supplementary condition.
Note that on-shall   last two relations in first  line are equivalent to  relations in second line,  but as we will see  it is useful to keep all relations.

These   Rarita-Schwinger  equations  can be combined  into  one equation  as linear combination of  Dirac equation and supplementary conditions
\begin{eqnarray}\label{eq:coN1}
\Big[  ( i {\hat{\partial}}  -   m  ) P_1  +   A  \pi_0   +  B  \pi  +  C   \pi_{\ast 0}  +   D   \pi_{0 \ast} \Big]^a{}_b  \psi^{b \alpha} (x)  =  0    \,  .  \qquad
\end{eqnarray}

Since   ${\hat{\partial}} $  commute with  projectors  $P_1,  \pi_0$  and  $\pi$  and anticommute with  $ \pi_{ \ast 0}$  and   $ \pi_{0 \ast}$  we can chose that all
coefficients  can depend on both   ${\hat{\partial}} $   and $m$  so that
$A =  a i {\hat{\partial}} +  \alpha m,  B =  b i {\hat{\partial}}  +  \beta  m$,     $  C  =  c i {\hat{\partial}} +  \gamma m$  and    $ D =  d i {\hat{\partial}} + \delta m $.

It is obvious that   (\ref{eq:coN1})   follows from   (\ref{eq:con11c}), but we will  show that the opposite is also true.
Since operator  $P_1$ annihilate  all other operators,   applying    operator $P_1$  on (\ref{eq:coN1})  we obtain
\begin{eqnarray}\label{eq:}
( i {\hat{\partial}}  -   m  ) P_1 \psi^a (x)  = 0  \,  .  \qquad
\end{eqnarray}

Then using relation between projectors and  applying $\pi_0, \, \pi, \, \pi_{\ast 0}$   and  $\pi_{0 \ast}$  on (\ref{eq:coN1})  we have
\begin{eqnarray}\label{eq:}
\Big(  A  \pi_0   +   D   \pi_{0 \ast} \Big)^a{}_b  \psi^{b \alpha} (x)  =  0    \,  ,   \qquad
\Big(    B  \pi  +  C   \pi_{\ast 0}   \Big)^a{}_b  \psi^{b \alpha} (x)  =  0    \,  ,   \qquad     \nonumber \\
\Big(  \tilde{A} \pi_{\ast 0}   +   \tilde{D}   \pi \Big)^a{}_b  \psi^{b \alpha} (x)  =  0    \,  ,   \qquad
\Big(   \tilde{ B}  \pi_{0 \ast}  + \tilde{ C}   \pi_0    \Big)^a{}_b  \psi^{b \alpha} (x)  =  0    \,  .   \qquad
\end{eqnarray}
Here $  \tilde{A} ({\hat{\partial}}, m ) =  A  ( -{\hat{\partial}}, m )  $, since ${\hat{\partial}} $  anticommute with $\pi_{0 \ast}$   and $\pi_{\ast 0}$.

So, for $A  \tilde{B}  - \tilde{C} D  \ne 0$ we obtain all conditions  (\ref{eq:con11c}),   which means that    from   (\ref{eq:coN1})   follows   (\ref{eq:con11c}), and that  these equations are equivalent.

\section{Regular form  of  spin-$\frac{1}{2}$  Rarita-Schwinger equation }
\setcounter{equation}{0}

As we have already discussed  all  operators   in  Eq.(\ref{eq:coN120})  contain singularities $\partial^{- 2}$.
So, we are going to   chose coefficients $A, B, C $  and $D$   in such a way   that equation  (\ref{eq:coN120}) becomes  regular.

\subsection{Regular form  of Eq. I}

We can separate kinetic and mass terms of the  first equation   (\ref{eq:coN120})
\begin{eqnarray}\label{eq:}
  K^a{}_b  =   i {\hat{\partial}}   \Big(   \pi_0  +   a  P_1    +  b  \pi   \Big)^a{}_b        \,  ,  \qquad
 M^a{}_b  =  m \Big(  -  \pi_0  +   \alpha  P_1    +  \beta  \pi   \Big)^a{}_b         \,  .  \qquad
\end{eqnarray}
Using  relation   $P_1 =   1 -   \pi_0 - \pi$  we obtain
\begin{eqnarray}\label{eq:}
  K^a{}_b  =    i {\hat{\partial}}  \Big[ a + (1 - a)   \pi_0     + ( b - a )  \pi   \Big]^a{}_b        \,  ,  \qquad
 M^a{}_b  =  m \Big[ \alpha     - (1 + \alpha)  \pi_0     + ( \beta - \alpha )  \pi   \Big]^a{}_b         \,  .  \qquad
\end{eqnarray}

Singular parts of kinetic  and mass terms are
\begin{eqnarray}\label{eq:}
 ( K_S)^a{}_b  =    \Big[ 1 - a    +  \frac{2}{3} ( b - a )   \Big]    ( i {\hat{\partial}} \pi_0)^a{}_b        \,  ,  \qquad      \nonumber \\
( M_S)^a{}_b =  m \Big[  \frac{\beta - 4 \alpha - 3 }{3}  (\pi_0 )^a{}_b     +   \frac{\beta - \alpha}{3}  \Big(  -  \gamma^a  (\rho_0)_b - \rho_0^a \gamma_b   \Big)   \Big]    \,  .  \qquad
\end{eqnarray}
So, to eliminate  singular  parts we  must require
\begin{eqnarray}\label{eq:}
  1 - a    +  \frac{2}{3} ( b - a )   = 0      \,  ,  \qquad
 \beta - 4 \alpha - 3  = 0  \, ,  \qquad       \beta - \alpha  = 0     \,  .  \qquad
\end{eqnarray}
or for $ 1- a = 2 \kappa$
\begin{eqnarray}\label{eq:}
  b  = 1  - 5 \kappa      \,  ,  \qquad    b - a  = - 3 \kappa  \, ,    \qquad     \beta = \alpha  =  - 1     \,  .  \qquad
\end{eqnarray}

Then regular parts are
\begin{eqnarray}\label{eq:}
 ( K_R)^a{}_b  =    i {\hat{\partial}}  \Big[ 1 - 2 \kappa + 2 \kappa  \pi_0     - 3 \kappa  \pi   \Big]^a{}_b        \,  ,  \qquad
 (M_R )^a{}_b  =   - m  \delta^\alpha_\beta     \,  .  \qquad
\end{eqnarray}

So,   complete regular spin-$\frac{1}{2}$  Rarita-Schwinger  equation   $( K_R  + M_R)^a{}_b  \psi^b  = 0$   in terms of  projectors   is
\begin{eqnarray}\label{eq:}
 \Big[    i {\hat{\partial}} - m   +   \kappa  i {\hat{\partial}}  \Big(  - 2  + 2   \pi_0     - 3   \pi  \Big)  \Big]^a{}_b   \psi^b  =  0     \,  .  \qquad
\end{eqnarray}
For $\kappa = 0 $  this is standard Dirac equation.

Applying operator $\pi_0$  we obtain we obtain  Dirac equation for representation $\psi_{0 1}^a  = ( \pi_0 )^a{}_b   \psi^b$
\begin{eqnarray}\label{eq:}
 \Big(    i {\hat{\partial}} - m   \Big)   ( \pi_0 )^a{}_b   \psi^b  =  0     \,  .  \qquad
\end{eqnarray}

\subsection{Regular form  of Eq. II}

We can separate kinetic and mass terms of the  second  equation   (\ref{eq:coN120})
\begin{eqnarray}\label{eq:}
  K^a{}_b  =   i {\hat{\partial}}   \Big(   \pi  +   c  P_1    +  d  \pi_0   \Big)^a{}_b        \,  ,  \qquad
 M^a{}_b  =  m \Big(  -  \pi  +   \gamma  P_1    +  \delta  \pi_0   \Big)^a{}_b         \,  ,  \qquad
\end{eqnarray}
and  using   expression   $P_1 =   1 -   \pi_0 - \pi$  we obtain
\begin{eqnarray}\label{eq:}
  K^a{}_b  =    i {\hat{\partial}}  \Big[ c + (d - c)   \pi_0     + ( 1 - c )  \pi   \Big]^a{}_b        \,  ,  \qquad
 M^a{}_b  =  m \Big[ \gamma  + (\delta  -  \gamma )  \pi_0     - ( 1 + \gamma )  \pi   \Big]^a{}_b         \,  .  \qquad
\end{eqnarray}

Singular parts of kinetic  and mass terms are
\begin{eqnarray}\label{eq:}
 ( K_S)^a{}_b  =    \Big[ d - c   +  \frac{2}{3} (1  - c )   \Big]    ( i {\hat{\partial}} \pi_0)^a{}_b        \,  ,  \qquad      \nonumber \\
( M_S)^a{}_b  =  m \Big[  \frac{ 3  \delta   - 4 \gamma - 1 }{3}  (\pi_0 )^a{}_b    -   \frac{1 +  \gamma }{3}  \Big(  -  \gamma^a  (\rho_0)_b - \rho_0^a \gamma_b   \Big)   \Big]    \,  .  \qquad
\end{eqnarray}
So, to eliminate  singular  parts we  must require
\begin{eqnarray}\label{eq:}
  d - c    +  \frac{2}{3} ( 1 - c  )   = 0      \,  ,  \qquad
3  \delta   - 4 \gamma - 1   = 0  \, ,  \qquad      \gamma  =  - 1     \,  ,  \qquad
\end{eqnarray}
or for $ 1- c = 3 \kappa$
\begin{eqnarray}\label{eq:}
  d - c  =  - 2 \kappa      \,  ,     \qquad     \gamma  =  \delta  =  - 1     \,  .  \qquad
\end{eqnarray}

Then regular parts are
\begin{eqnarray}\label{eq:}
 ( K_R)^a{}_b  =    i {\hat{\partial}}  \Big[1    +  \kappa  \Big(  - 3  - 2   \pi_0   + 3   \pi  \Big)   \Big]^a{}_b        \,  ,  \qquad
 (M_R )^a{}_b  =   - m  \delta^\alpha_\beta     \,  .  \qquad
\end{eqnarray}

So,   complete regular  spin-$\frac{1}{2}$  Rarita-Schwinger  equation   $( K_R  + M_R)^a{}_b  \psi^b  = 0$  in terms of  projectors   takes a form
\begin{eqnarray}\label{eq:}
 \Big[    i {\hat{\partial}} - m   +   \kappa  i {\hat{\partial}}  \Big(  - 3  - 2   \pi_0   + 3   \pi  \Big)  \Big]^a{}_b   \psi^b  =  0     \,  .  \qquad
\end{eqnarray}
For $\kappa = 0 $  this is standard Dirac equation.

Applying operator $\pi$  we obtain we obtain  Dirac equation for representation $\psi_{0 2}^a  = ( \pi )^a{}_b   \psi^b$
\begin{eqnarray}\label{eq:}
 \Big(    i {\hat{\partial}} - m   \Big)   ( \pi )^a{}_b   \psi^b  =  0     \,  .  \qquad
\end{eqnarray}

Consequently,  spin-$\frac{1}{2}$  parts of vector-spinor equation describe two Dirac spinors.

\section{Regular form  of  spin-$\frac{3}{2}$  Rarita-Schwinger equation }
\setcounter{equation}{0}

All  operators   in  Eq.(\ref{eq:coN1})  contain singularities $\partial^{- 2}$.   It means that we should  chose coefficients $A, B, C $  and $D$ so that equation  (\ref{eq:coN1}) becomes  regular.
Using  relation  $P_1 =   1 -   \pi_0 - \pi$  we can rewrite   (\ref{eq:coN1})  in the  form
\begin{eqnarray}\label{eq:coN}
\Big[  ( i {\hat{\partial}}  -   m  )  (1 -   \pi_0 - \pi  )   +   A  \pi_0   +  B  \pi  +  C   \pi_{\ast 0}  +   D   \pi_{0 \ast} \Big]^a{}_b  \psi^{b \alpha} (x)  =  0    \,  .  \qquad
\end{eqnarray}
Both kinetic  and mass  terms
\begin{eqnarray}\label{eq:coNp}
  K^a{}_b  =  i {\hat{\partial}} \Big[1  +  (- 1 + a)   \pi_0   +  (-1 + b)  \pi  +  c   \pi_{\ast 0}  +   d   \pi_{0 \ast} \Big]^a{}_b     \,  ,  \qquad  \nonumber \\
 M^a{}_b  =   m  \Big[- 1 +   ( 1 + \alpha)  \pi_0   +   ( 1 + \beta)  \pi  +  \gamma   \pi_{\ast 0}  +   \delta   \pi_{0 \ast} \Big]^a{}_b    \,  ,  \qquad
\end{eqnarray}
should be  regular  separately

In order to distinguish  singular terms we need explicit expressions  for variables
\begin{eqnarray}\label{eq:}
  (\rho_0 )^{a \alpha}{}_\gamma     =   (P_L)^a{}_b   ( \gamma^b )^\alpha{}_\beta    = \frac{\partial^a \hat{\partial}}{\partial^2}   \, ,  \qquad
    (\rho )^{a \alpha}{}_\gamma     =    ( \gamma^a )^\alpha{}_\beta -  (\rho_0 )^{a \alpha}{}_\gamma     \, ,    \qquad
\end{eqnarray}
as well as  for projectors
\begin{eqnarray}\label{eq:pi0pip}
(\pi_0)^a{}_b  (p)     =    \frac{\partial^a  \partial_b}{\partial^2}      \, ,      \qquad
(\pi_1)^a{}_b  (p)   =  \delta^a_b -  (\pi_0)^a{}_b        \, ,  \qquad \qquad \qquad                \nonumber \\
 (\pi)^a{}_b   (p) =  \frac{1}{3}  \Big( \gamma^a   \gamma_b  -  \gamma^a  (\rho_0)_b - \rho_0^a \gamma_b   +  (\pi_0)^a{}_b   \Big)     \, ,  \qquad  \qquad \qquad \qquad    \nonumber \\
(\pi_{0 \ast})^a{}_b (p)   =   \frac{1}{\sqrt{3}}    \Big( \rho_0^a \gamma_b -  (\pi_0)^a{}_b   \Big)        \, ,  \qquad
  (\pi_{\ast 0})^a{}_b    (p)   =   \frac{1}{\sqrt{3}}    \Big( \gamma^a   ( \rho_0)_b  -  (\pi_0)^a{}_b   \Big)        \, .  \qquad
\end{eqnarray}

Therefore,  basic singular terms are
\begin{eqnarray}\label{eq:}
\pi_0,  \quad  \gamma^a  (\rho_0)_b,   \quad   \rho_0^a \gamma_b                   \,  .
\end{eqnarray}
If we multiply them with $\hat{\partial}$  then singular terms are
\begin{eqnarray}\label{eq:}
\hat{\partial}  \pi_0  \, ,  \quad
\hat{\partial}  \gamma^a  (\rho_0)_b    =  2   \hat{\partial}  \pi_0   -   \gamma^a \partial_b       \, ,
\end{eqnarray}
so that in kinetic part  only singular expression  is $\hat{\partial}  \pi_0 $.

\subsection{Regular form  of kinetic term  }

Separating singular parts of kinetic term    (\ref{eq:coNp})   we have
\begin{eqnarray}\label{eq:coNp1}
( K_S )^a{}_b  =   \Big( - 1 + a  + \frac{( - 2 + 1 )( - 1 + b) }{3}  +  \frac{2 c}{\sqrt{3}}  - \frac{c + d}{\sqrt{3}}   \Big)  \Big( i \hat{\partial}  \pi_0  \Big)^a{}_b \,  ,  \qquad
\end{eqnarray}
 so that condition that this part is free of singularities is
\begin{eqnarray}\label{eq:}
  a - 1 - \frac{ b - 1}{3}  +  \frac{c -  d }{\sqrt{3}}    = 0           \,  .  \qquad
\end{eqnarray}

Consequently, regular part is
\begin{eqnarray}\label{eq:}
( K_R )^a{}_b   =  i {\hat{\partial}} \Big[1     +    \frac{ b - 1}{3} ( \pi_0 +  3 \pi  )
  +    \frac{ c}{\sqrt{3}}   ( \sqrt{3} \pi_{\ast 0}  - \pi_0  ) +    \frac{ d }{\sqrt{3}}  ( \sqrt{3} \pi_{0 \ast}  + \pi_0  ) \Big]^a{}_b  \,  .  \qquad
\end{eqnarray}

It means that expressions in terms of projectors corresponding to  each arbitrary choice of  parameters  must be regular. Let us calculate them  explicitly.
\begin{eqnarray}\label{eq:}
 {\hat{\partial}}  ( \pi_0 +  3 \pi  )    =   -   \gamma^a  {\hat{\partial}}   \gamma_b  +   \partial^a  \gamma_b   +   \gamma^a  \partial_b        \,  ,  \qquad     \nonumber \\
 {\hat{\partial}}   ( \sqrt{3} \pi_{\ast 0}  - \pi_0  )   = -   \gamma^a  \partial_b    \,  ,  \qquad
 {\hat{\partial}}   ( \sqrt{3} \pi_{0 \ast}  + \pi_0  )  =    \partial^a \gamma_b    \,  .  \qquad
\end{eqnarray}

So, explicit expression  for regular part is
\begin{eqnarray}\label{eq:KReex}
( K_R )^a{}_b =   i  \Big[  \delta^a_b {\hat{\partial}}   -    \frac{ b - 1}{3} \gamma^a  {\hat{\partial}}   \gamma_b
+  \Big(    \frac{ b - 1}{3} -  \frac{ c}{\sqrt{3}}  \Big)   \gamma^a  \partial_b     +   \Big(    \frac{ b - 1}{3}  + \frac{ d }{\sqrt{3}}  \Big)   \partial^a \gamma_b      \Big]          \,  ,  \qquad
\end{eqnarray}

\subsection{Regular form of  mass   term  }

To eliminate  singular parts    $\pi_0$,  $\gamma^a  (\rho_0)_b$   and  $\rho_0^a \gamma_b $     from  equation  (\ref{eq:coNp})   the conditions
\begin{eqnarray}\label{eq:}
  1 +   \alpha    +  \frac{1  +  \beta}{3}   - \frac{1}{\sqrt{3}} (  \gamma   +   \delta )   =  0    \,  ,  \qquad   \nonumber \\
  - \frac{1  + \beta}{3}   + \frac{1}{\sqrt{3}}  \,  \gamma   =  0    \,  ,  \qquad
 - \frac{1 + \beta}{3}   + \frac{1}{\sqrt{3}}  \,  \delta   =  0    \,  , \qquad
\end{eqnarray}
must be satisfy. The solution is
\begin{eqnarray}\label{eq:}
  1 +   \alpha    =  \frac{1  +  \beta}{3}       \,  ,  \qquad
 \gamma  = \delta   =   \frac{1 + \beta}{\sqrt{3}}    \,  ,  \qquad
\end{eqnarray}
which for $1 + \beta = \kappa$  produces regular part of mass term
\begin{eqnarray}\label{eq:coNmpf0}
( M_R )^a{}_b  =  - m  +  \frac{\kappa}{3} m \Big[     \pi_0   + 3  \pi +  \sqrt{3}   (   \pi_{\ast 0}  +  \pi_{0 \ast}  ) \Big]^a{}_b    \,  .  \qquad
\end{eqnarray}

Using  relation
\begin{eqnarray}\label{eq:}
\Big( \pi_0  +  3 \pi   +   \sqrt{3}  ( \pi_{0 \ast}  + \pi_{ \ast 0}   )   \Big)^a{}_b    = ( \gamma^a \gamma_b )^\alpha{}_\beta  \, .
\end{eqnarray}
we obtain
\begin{eqnarray}\label{eq:coNmpf0}
( M_R )^a{}_b  =   \frac{\kappa}{6} m   [ \gamma^a,  \gamma_b ]   +  \Big( \frac{\kappa}{3}  - 1  \Big)  m   \delta^a_b    \,  .   \qquad
\end{eqnarray}

\subsection{Regular form of  Rarita-Schwinger  equation }

The   complete regular   Rarita-Schwinger  equation   $( K_R  + M_R)^a{}_b  \psi^b  = 0$   in terms of  projectors  have  a form
\begin{eqnarray}\label{eq:cRSP}
 \Big[  i {\hat{\partial}} - m      +    \frac{ b - 1 }{3}  i {\hat{\partial}}  ( \pi_0 +  3 \pi  )   + \frac{ c}{\sqrt{3}}  i {\hat{\partial}}  ( \sqrt{3} \pi_{\ast 0}  - \pi_0  )
 +     \frac{ d }{\sqrt{3}} i {\hat{\partial}} ( \sqrt{3} \pi_{0 \ast}  + \pi_0  )                        \nonumber \\
+  \frac{\kappa}{3} m \Big(     \pi_0   + 3  \pi +  \sqrt{3}   (   \pi_{\ast 0}  +  \pi_{0 \ast}  )   \Big)    \Big]^a{}_b  \psi^b  =  0    \,  ,  \qquad
\end{eqnarray}
and explicitly
\begin{eqnarray}\label{eq:cRSe}
      \delta^a_b ( i {\hat{\partial}} - m )  -    \frac{ b - 1}{3} \gamma^a  {\hat{\partial}}   \gamma_b
+  \Big(    \frac{ b - 1}{3} -  \frac{ c}{\sqrt{3}}  \Big)   \gamma^a i \partial_b     +   \Big(    \frac{ b - 1}{3}  + \frac{ d }{\sqrt{3}}  \Big)  i  \partial^a \gamma_b
+  \frac{\kappa}{6} m   [ \gamma^a,  \gamma_b ]   +   \frac{\kappa}{3}  m   \delta^a_b   = 0         \,  .  \qquad
\end{eqnarray}

It is useful  to  rewrite regular  Rarita-Schwinger  equation in terms of spin-$\frac{3}{2}$  projector $P_1$. So, using relation
\begin{eqnarray}\label{eq:}
 \Big( P_1 -  2 \pi  \Big)^a{}_b  =   \Big( 1  -   \pi_0  - 3  \pi    \Big)^a{}_b     \,  ,  \qquad
\end{eqnarray}
we obtain
\begin{eqnarray}\label{eq:cRSP1}
 \Big[ ( i {\hat{\partial}} - m )  ( P_1 -  2  \pi)    +    \frac{ b + 2 }{3}  i {\hat{\partial}}  ( \pi_0 +  3 \pi  )   + \frac{ c}{\sqrt{3}}  i {\hat{\partial}}  ( \sqrt{3} \pi_{\ast 0}  - \pi_0  )
 +     \frac{ d }{\sqrt{3}} i {\hat{\partial}} ( \sqrt{3} \pi_{0 \ast}  + \pi_0  )                        \nonumber \\
+  \frac{\kappa - 3}{3} m \Big(     \pi_0   + 3  \pi \Big) +     \frac{\kappa}{ \sqrt{3}} m \Big(    \pi_{\ast 0}  +  \pi_{0 \ast}   \Big)    \Big]^a{}_b  \psi^b  =  0    \,  .  \qquad
\end{eqnarray}

The equations  (\ref{eq:cRSP}),  (\ref{eq:cRSe})  and   (\ref{eq:cRSP1})  are  equivalent and  they describe the  most general forms  of regular  Rarita-Schwinger  equation.  They have  four arbitrary parameters $b, c, d$  and $\kappa$.

\subsection{Symmetric form of the action    }

We are going to check  consistency  condition,  similar to what is required  in Ref.\cite{N}.  This is statement   that variation  of the action  with respect to $ \psi^b$  produces the same equation as
variation with respect to  ${\bar \psi}^a$.

We can rewrite the equation (\ref{eq:cRSP1})  in the form
\begin{eqnarray}\label{eq:cRSP1n}
 \Big\{ ( i {\hat{\partial}} - m )   P_1    +   \Big[  \Big(  \frac{ b + 2 }{3} -  \frac{ c  - d}{\sqrt{3}} \Big)    i {\hat{\partial}}  +   \frac{\kappa - 3}{3} m     \Big]  \pi_0
  +   \Big[    b  i {\hat{\partial}}  +  ( \kappa - 1 )  m     \Big]  \pi           \nonumber \\
 +  \Big( c  i {\hat{\partial}}   +  \frac{\kappa}{ \sqrt{3}} m   \Big) \pi_{\ast 0}
 +     \Big( d  i {\hat{\partial}}   +  \frac{\kappa}{ \sqrt{3}} m   \Big)   \pi_{0 \ast}      \Big\}^a{}_b  \psi^b  =  0    \,  .  \qquad
\end{eqnarray}
This  equation   can be obtained varying  the action
\begin{eqnarray}\label{eq:}
 S_{RS}  =  \int  d^4 x  {\bar \psi}^a   {\cal L}_{a b}  \psi^b
 \equiv \int  d^4 x  {\bar \psi}^a   \Big\{ ( i {\hat{\partial}} - m ) P_1 + \Big[  \Big(  \frac{ b + 2 }{3} -  \frac{ c  - d}{\sqrt{3}} \Big) i {\hat{\partial}}  +   \frac{\kappa - 3}{3} m     \Big]  \pi_0  \nonumber \\
  +   \Big[    b  i {\hat{\partial}}  +  ( \kappa - 1 )  m     \Big]  \pi
 +  \Big( c  i {\hat{\partial}}   +  \frac{\kappa}{ \sqrt{3}} m   \Big) \pi_{\ast 0}
 +     \Big( d  i {\hat{\partial}}   +  \frac{\kappa}{ \sqrt{3}} m   \Big)   \pi_{0 \ast}      \Big\}_{a b}   \psi^b    \, ,   \qquad  \qquad
\end{eqnarray}
with respect to ${\bar \psi}^a  $.

The   variation with respect to $ \psi^b  $  should   produce the same  field equation. Taking into account that Lagrangian must be real   we have
\begin{eqnarray}\label{eq:}
  \int  d^4 x  {\bar \psi}^a   {\cal L}_{a b} \delta \psi^b   =     \int  d^4 x  \delta {\bar \psi}^b \gamma^0   {\cal L}^\dagger_{a b}  \gamma^0  \psi^a     \, .  \qquad
   \, ,   \qquad  \qquad
\end{eqnarray}
After variation with respect to $  {\bar \psi}^b $ we obtain equation   $\gamma^0   {\cal L}^\dagger_{a b}  \gamma^0  \psi^a = 0$.
In order  that  this equation is    equivalent  to initial one we must require
\begin{eqnarray}\label{eq:g0g0}
\gamma^0   {\cal L}^\dagger_{a b}  \gamma^0 =  {\cal L}_{b a}  \, .
\end{eqnarray}

Operators $P_1, \pi_0, \pi$ satisfy relations
\begin{eqnarray}\label{eq:}
\gamma^0   (P_1^\dagger)_{a b}  \gamma^0 =  (P_1)_{b a}  \, ,  \qquad   \gamma^0   (\pi_0^\dagger)_{a b}  \gamma^0 =  (\pi_0)_{b a}  \, ,  \qquad     \gamma^0   \pi^\dagger_{a b}  \gamma^0 =  \pi_{b a}  \, ,  \qquad
\end{eqnarray}
while  for  $\pi_{0 \ast } $,  $\pi_{ \ast 0} $  and   $ i {\hat{\partial}}$ we have
\begin{eqnarray}\label{eq:}
\gamma^0   (\pi_{\ast 0}^\dagger)_{a b}  \gamma^0 =    (\pi_{0 \ast })_{b a}     \, ,  \qquad    \gamma^0   (\pi_{0 \ast }^\dagger)_{a b}  \gamma^0 =    (\pi_{ \ast 0})_{b a}     \, ,  \qquad
\gamma^0  (i {\hat{\partial}} )^\dagger \gamma^0 =  - i {\hat{\partial}}   \, ,   \qquad
\end{eqnarray}
But we must perform one partial integration  which destroys the minus sign  in last relation.
We  should  also  take into account that  ${\hat{\partial}}  $  anticommute with operators $\pi_{\ast 0}$  and  $\pi_{0 \ast }$  and commute with other operators.
So, the terms   $c  i {\hat{\partial}} \pi_{\ast 0} $     and   $d  i {\hat{\partial}} \pi_{0 \ast } $  change the sign and exchange the places and  according to   (\ref{eq:g0g0})   we should take $c  = - d$.

Then we obtain the  most general form of the action which depend on three parameters  $b, c$   and $\kappa$
\begin{eqnarray}\label{eq:RS3p}
 S_{RS}  =  \int  d^4 x  {\bar \psi}^a   {\cal L}_{a b}  \psi^b
  \equiv \int  d^4 x  {\bar \psi}^a   \Big\{ ( i {\hat{\partial}} - m ) P_1 + \Big[  \Big(  \frac{ b + 2 }{3} -\frac{2  c}{\sqrt{3}} \Big)    i {\hat{\partial}}  +   \frac{\kappa - 3}{3} m     \Big]  \pi_0  \nonumber \\
  +   \Big[    b  i {\hat{\partial}}  +  ( \kappa - 1 )  m     \Big]  \pi    +  \Big( c  i {\hat{\partial}}   +  \frac{\kappa}{ \sqrt{3}} m   \Big) \pi_{\ast 0}
 +     \Big( - c  i {\hat{\partial}}   +  \frac{\kappa}{ \sqrt{3}} m   \Big)   \pi_{0 \ast}      \Big\}_{a b}   \psi^b    \, .   \qquad  \qquad
\end{eqnarray}
In components  it takes the form
\begin{eqnarray}\label{eq:cRSe1}
  S_{RS}  =
\int  d^4 x  {\bar \psi}^a   \Big[  \eta_{a b} ( i {\hat{\partial}} - m )  -    \frac{ b - 1}{3} \gamma_a  {\hat{\partial}}   \gamma_b
+  \Big(    \frac{ b - 1}{3} -  \frac{ c}{\sqrt{3}}  \Big)   \Big(  \gamma_a i \partial_b     +   i  \partial_a \gamma_b  \Big)                     \nonumber \\
+  \frac{\kappa}{6} m   [ \gamma_a,  \gamma_b ]   +   \frac{\kappa}{3}  m    \eta_{a b}   \Big]   \psi^b           \,  .  \qquad
\end{eqnarray}

There is a set of parameters
\begin{eqnarray}\label{eq:}
  b = -  \frac{1}{8}   \,  ,  \qquad     c =   \frac{\sqrt{3}}{8}   \,  ,  \qquad   \kappa  =  \frac{3}{4}
\end{eqnarray}
for which above   expressions  are   incorrect   since  relation   $A  \tilde{B}  = \tilde{C} D  $ is satisfy.    In that case   relations   (\ref{eq:con11c})    and  (\ref{eq:coN1})    are  not  equivalent  and
we are not able to  combined    Dirac equation and supplementary conditions into  one equation.

\subsection{Particular  choice of parameters leads to Nieuwenhuizen and    Rarita-Schwinger form }

In particular case  for  $b=  -2, \, c  = 0$  and $\kappa = 3$  from      (\ref{eq:RS3p})   we have
\begin{eqnarray}\label{eq:RSN}
 \Big[  i {\hat{\partial}}   ( P_1 -  2  \pi)    - m \Big(   P_1 -  2  \pi -   \sqrt{3}  \Big( \pi_{\ast 0}  +  \pi_{0 \ast}   \Big)  \Big)   \Big]^a{}_b  \psi^b  =  0    \,  .  \qquad
\end{eqnarray}
This is exactly equation obtained in Ref.\cite{N}, where   connections  between   Nieuwenhuizen and our  notation  is
\begin{eqnarray}\label{eq:IrR1rs10fn}
\Pi^{\frac{3}{2}}_{a b}  =  ( P_1  )_{a b}   \, ,  \qquad
\Big( \Pi^{\frac{1}{2}}_{2 2} \Big)_{a b}   =  ( \pi_0 )_{a b}   \, ,  \qquad     \Big( \Pi^{\frac{1}{2}}_{1 1} \Big)_{a b}  =   ( \pi )_{a b}     \, ,   \qquad   \nonumber \\
\Big( \Pi^{\frac{1}{2}}_{1 2} \Big)_{a b}   =    ( \pi_{ \ast 0} )_{a b}   \, ,  \qquad   \Big( \Pi^{\frac{1}{2}}_{2 1} \Big)_{a b}   =    ( \pi_{0 \ast } )_{a b}  \, .  \qquad
\end{eqnarray}

With the help of the relations
\begin{eqnarray}\label{eq:Kint1}
i  {\hat \partial}  ( P_1 -  2 \pi  )^{a d}   =  \varepsilon^{a b c d} \gamma^5  \gamma_b \partial_c     \,  ,
\end{eqnarray}
and
\begin{eqnarray}\label{eq:masst1}
\Big(   P_1  -   2 \pi    -  \sqrt{3}  ( \pi_{0 \ast}  + \pi_{ \ast 0}   )   \Big)^a{}_b    =  - \frac{1}{2}  [ \gamma^a,  \gamma_b   ]  \,  ,
\end{eqnarray}
we can rewrite    (\ref{eq:RSN})     in the form
\begin{eqnarray}\label{eq:}
   \varepsilon^{a b c d} \gamma^5  \gamma_b \partial_c  \psi_d (x)    +     \frac{m}{2}  [ \gamma^a,  \gamma_b   ] \psi^b (x)    =  0    \,  .    \qquad
\end{eqnarray}
This is Rarita-Schwinger  equation  \cite{RS}  which can be obtained from the Lagrangian
\begin{eqnarray}\label{eq:}
{\cal L} = - \frac{1}{2} {\bar \psi}_a \Big(  \varepsilon^{a b c d}  \gamma^5 \gamma_b \partial_c  -  i m  \sigma^{a d}  \Big)  \psi_d  \, ,   \qquad    \sigma_{a b} = \frac{i}{2} \, [\gamma_a , \gamma_b]   \, .
\end{eqnarray}

\section{Conclusion   }
\setcounter{equation}{0}

The method used in  previous two articles  \cite{BS1, BS2}  shows that all theories with same symmetry group will look the same at sufficiently  low energy.
This confirms Weinberg suggestion relating to Poincare group \cite{W}.

In Ref. \cite{BS2}   we have shown  that Poincare  group  completely determine the  the spectrum   of massless particles  and corresponding free field  equations.
We have given an examples of  massless  vector,  second and fourth rang tensor  fields and obtain Maxwell equations  and Einstein equation in weak field approximation.

In massive case in Ref.\cite{BS1}   we described scalar, Dirac  and vector fields. Here we continue this consideration and  derived  massive Rarita-Schwinger  action
for  spin-$\frac{3}{2}$  field,  originally  introduced in Re.\cite{RS}.
In order to simplify calculations we introduced   new  approach based on  projection operators.  In particular case  Rarita-Schwinger  projectors can be explain in terms of vector and Dirac projectors.
Unlike massless case here in massive case there is mixed term of basic vector and spinor  projectors.

Consequently,   we obtained  the most general action  for regular spin-$\frac{3}{2}$  vector-spinor field   (\ref{eq:RS3p}). It depend on three parameters. In particular case we can obtain
Nieuwenhuizen and    Rarita-Schwinger form of the action.
Let us stress that  the most  important contribution  of the paper is  set of   principle field equations
(\ref{eq:BS}).   In  non-trivial example  of  Rarita-Schwinger equation  we confirmed our claim from  Ref.\cite{BS1}  that we can unify free  massive  field  equations for arbitrary spin
in two principle  equations.

\end{document}